\documentclass[%
 reprint,
superscriptaddress,
 amsmath,amssymb,
 aps,
]{revtex4-1}

\usepackage{aas_macros}
\usepackage{amsfonts}
\usepackage{amsmath}
\usepackage{amssymb}
\usepackage{upgreek}
\usepackage{bm}
\usepackage{dcolumn}
\usepackage{epsfig}
\usepackage{graphicx}
\usepackage{graphics}
\usepackage{placeins}
\usepackage{subfigure}

\usepackage{color}

\newcommand{\dif}{{\rm d}}
\newcommand{\degree}{\ensuremath{^\circ}}

\def\beq{\begin{equation}}
\def\eeq{\end{equation}}
\def\bi{\begin{itemize}}
\def\ei{\end{itemize}}
\def\ben{\begin{enumerate}}
\def\een{\end{enumerate}}
\def\bea{\begin{eqnarray}}
\def\eea{\end{eqnarray}}

\DeclareMathOperator{\var}{var}

\begin{document}

\title{Understanding $\Omega_\mathrm{gw}(f)$ in Gravitational Wave Experiments}

\author{Chiara M.~F.~Mingarelli}

\email{chiara.mingarelli@yale.edu}
\affiliation{Department of Physics, Yale University, New Haven, CT 06520, USA}
\affiliation{Department of Physics, University of Connecticut, 196 Auditorium Road, U-3046, Storrs, CT 06269-3046, USA}
\affiliation{Center for Computational Astrophysics, Flatiron Institute, 162 Fifth Ave, New York, NY 10010, USA }

\author{Stephen~R.~Taylor}
\affiliation{Department of Physics \& Astronomy, Vanderbilt University,
2301 Vanderbilt Place, Nashville, TN 37235, USA}
\affiliation{TAPIR, California Institute of Technology,
1200 E California Blvd., Pasadena, CA 91125, USA }

\author{B.~S. Sathyaprakash}
\affiliation{Department of Physics and Department of Astronomy and Astrophysics,
The Pennsylvania State University, University Park, PA 16802, USA}
\affiliation{School of Physics and Astronomy, Cardiff University, Cardiff, CF24 3AA, UK}

\author{Will M. Farr}
\affiliation{Center for Computational Astrophysics, Flatiron Institute, 162 Fifth Ave, New York, NY 10010, USA }
\affiliation{Department of Physics and Astronomy, Stony Brook University, Stony Brook NY 11794, USA}
\date{\today}

\begin{abstract}
In this paper we provide a comprehensive derivation of the energy density in the stochastic gravitational-wave background $\Omega_\mathrm{gw}(f)$, and show how this quantity is measured in ground-based detectors such as LIGO, the space-based Laser Interferometer Space Antenna (LISA), and Pulsar Timing Arrays. By definition $\Omega_\mathrm{gw}(f) \propto S_h(f)$ -- the power spectral density (PSD) of the Fourier modes of the gravitational-wave background. However, this is often confused with the PSD of the strain signal, which we call $S_\mathrm{gw}(f)$, and is a detector-dependent quantity. This has led to confusing definitions of $\Omega_\mathrm{gw}(f)$ in the literature which differ by factors of up to 5 when written in a detector-dependent way. In addition to clarifying this confusion, equations presented in this paper facilitate easy comparison of results from different detector groups, and how to convert from one measure of the strength of the background to another.
\end{abstract}

\pacs{04.80.Nn, 95.55.Ym, 98.80.-k, 04.30.-w}

\maketitle

\section{Introduction}
\begin{figure*}[ht]
	\centering
	\includegraphics[width=\linewidth]{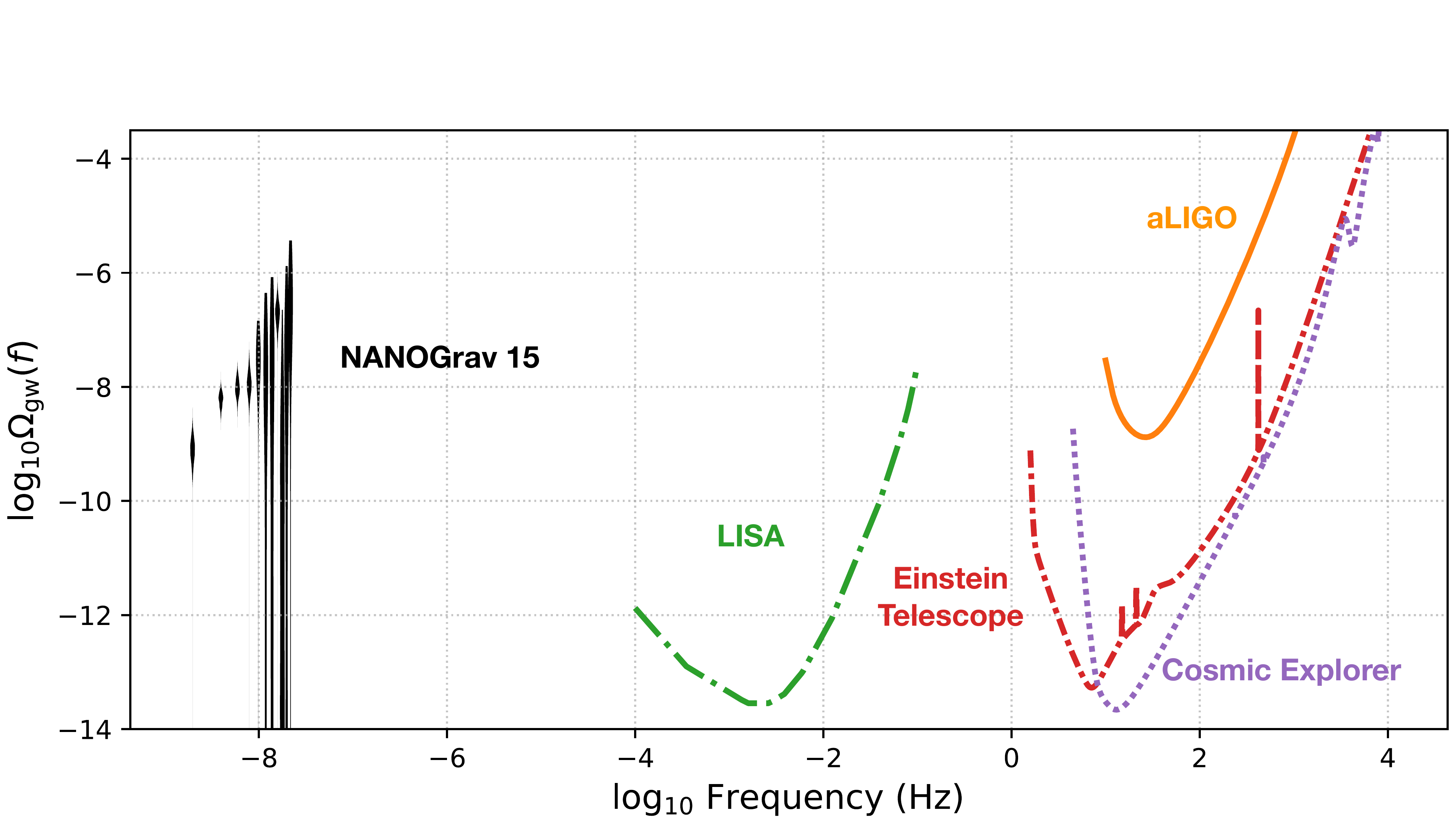}
	\caption[The Gravitational-Wave Landscape]{The current and future gravitational-wave landscape in terms of $\Omega_\mathrm{gw}(f)$, assuming $H_0=68$~km/s/Mpc \cite{collaboration2018planck}. Solid lines are experiments which are currently operational. For PTAs we show the measurement of the low-frequency gravitational-wave background from NANOGrav's 15 year data, and for ground-based interferometers we show LIGO. Dashed and dot-dashed lines are future experiments, including LISA \citep{sb09}, Einstein Telescope (ET) \cite{sba12, ET}, and Cosmic Explorer \citep{Evans:2016mbw}, assuming 2 years of data collection.  PTA data are from the NANOGrav 15 yr data release~\cite{NG15-psr}. 
    }
	\label{fig:gw_spec}
\end{figure*}

The new era of gravitational-wave (GW) astronomy arrived with a chirp from a binary black hole merger detected by the Laser Interferometer Gravitational-wave Observatory, (LIGO)  \cite{AbbottEtAl:2016, aLIGO:2016}. LIGO is the first GW experiment to directly detect gravitational radiation, however other GW detectors are poised to open up the full GW spectrum, Figure \ref{fig:gw_spec}. At the low frequency end, Pulsar Timing Arrays (PTAs), see e.g. \cite{Sazhin:1978, Detweiler1979, Hellings1983, 1990ApJ...361..300F, iptadr2, Mingarelli:2019,taylor2021nanohertz, Ming2025}, now have evidence for a GW background (GWB)~\cite{NG15-GWB, EPTA-GWB, PPTA-GWB, CPTA-GWB, miles2025meerkat}, likely from the cosmic merger history of supermassive black hole binaries (SMBHB)~\cite{Siemens2013, 2015MNRAS.451.2417R, tve+16, 2017MNRAS.471.4508K,caseyclyde2022}. Detections of nearby resolvable SMBHBs are expected to follow in the next decade, or sooner \cite{2015MNRAS.451.2417R, mls+2017, 2018MNRAS.477..964K,XMH21}, while the Laser Interferometer Space Antenna (LISA) \cite{Lisa:2017, Lisa:2019} will fill out the GW spectrum at millihertz frequencies in the 2030s.

Both astrophysical and cosmological sources are expected to contribute to a low-frequency GWB (see e.g. \cite{lms+16, 2018arXiv181108826B, 2019BAAS...51c.336T, 2019BAAS...51c.437S,NG15-newphysics}) and therefore a measurement of the amplitude of the GWB offers a new and exciting avenue to explore the evolution of the Universe.
A multitude of experiments have found evidence for, or set limits on, the amplitude of the GWB at different GW frequencies, thus putting limits on the GW energy density per logarithmic frequency. By dividing this quantity by the critical energy density to close the Universe, we write down $\Omega_\mathrm{gw}(f)$, Equation \eqref{eq:omegagw}. PTAs and LIGO can measure, or limit, $\Omega_\mathrm{gw}(f)$ at a reference frequency $f$, whereas constraints on $\Omega_\mathrm{gw}$ from Big Bang Nucleosynthesis and Cosmic Microwave Background experiments \cite{collaboration2018planck} report this value integrated over frequency. 

We derive $\Omega_\mathrm{gw}(f)$ comprehensively here since there has been some confusion in the field, e.g. \cite{SathyaSchutz:2009}, regarding its definition: the detector-independent quantity, $\Omega_\mathrm{gw}(f)\propto S_h(f)$, where $S_h(f)$ is the 1-sided power spectral density (PSD) of the Fourier modes of the GWB, versus the detector-dependent quantity,  $\Omega_\mathrm{gw}(f)\propto  S_\mathrm{gw}(f)$, where $S_\mathrm{gw}(f)$ is the measured PSD of the strain signal in the detector.

To clear up this confusion we show how $\Omega_\mathrm{gw}(f)$ maps on to GW detectors in terms of PSD of the strain signal $S_\mathrm{gw}(f)$ for the three main GW detectors which will be operating in the near future: LIGO, LISA, and PTAs. We show that for LIGO $S_h(f) = 5\,S_\mathrm{gw}(f)$, for LISA $S_h(f) = 5/(\sin^2\beta) S_\mathrm{gw}(f)$, where $\beta=60 \degree$ is the opening angle at a vertex, while for PTAs this is $3\,S_\mathrm{gw}(f)$.

While different approaches and variations of this derivation appear in the literature, e.g. \cite{AllenRomano:1999, RomanoCornish2017}, it is clear that a self-contained and complete derivation for the general definition of $\Omega_\mathrm{gw}(f)$ is still required -- written in a detector-independent way in terms of $S_h(f)$ -- and detector-specific expressions \cite{AasiEtAl:2014, AbbottEtAl:2007, AbbottEtAl:2004}, which are a function of $S_\mathrm{gw}(f)$.

The paper is laid out as follows: in Sec \ref{sec:derivation} we give a comprehensive derivation of $\Omega_\mathrm{gw}(f)$. In Sec \ref{sec:psdStrain}, we show how one obtains the PSD of the strain $S_\mathrm{gw}(f)$, and how to write  $\Omega_\mathrm{gw}(f)$ in terms this quantity for LIGO, LISA, and PTAs. Concluding remarks are given in Sec \ref{disc}. 
Unless otherwise specified, the work is carried out in natural units $c=G=1$.

\section{Derivation of $\Omega_\mathrm{gw}(f)$} \label{sec:derivation}
The starting point for the various ways of writing $\Omega_\mathrm{gw}(f)$ \cite{MTW, Maggiore:2000}, is
\beq
	\label{eq:omegagw}
	\Omega_\mathrm{gw}(f)=\frac{1}{\rho_c}\frac{d \rho_{\mathrm{gw}}}{d\log f} \, ,
\eeq
where $f$ is the frequency, $\rho_c=3H_0^2/8\pi$ is the critical energy density required to close the universe, $H_0 = 100~h$~km/s/Mpc is
the Hubble expansion rate, with $h$ the dimensionless Hubble parameter, and $\rho_{\mathrm{gw}}$ is the total energy density in GWs~\cite{AllenOttewill:1997, ar99}.

The stress-energy tensor of GWs is given by the Isaacson expression \cite{Isaacson:1968},
\begin{equation}
	T_{\mu\nu} = \frac{1}{32\pi}\langle\partial_{\mu}h_{ab}\partial_{\nu}h^{ab}\rangle\, ,
\end{equation}
where $\langle \rangle$ denote the average, and the energy density is given by the $00$ component. Therefore,
\begin{equation}
	\rho_\mathrm{gw} = \frac{1}{32\pi }\langle\dot{h}_{ab}\dot{h}^{ab}\rangle.
\end{equation}

We describe the metric perturbation in terms of a plane wave expansion, in the usual transverse traceless gauge: $h^{0\mu}=0,\, h^\mu_\mu=0$:

\begin{equation}
	h_{ab}(t,\vec{x}) = \!\! \!\sum_{A=+,\times}\int_{-\infty}^{\infty}\dif
	f\int_{S^2}\dif\hat{\Omega}\; h_A(f,\hat\Omega) e^{2\pi
  	if(t-\hat\Omega\cdot\vec{x})}e^A_{ab}(\hat\Omega) \, ,
\end{equation}
where $h_A(f,\hat\Omega)$ are the polarization amplitudes,  $\hat\Omega$ is the direction of propagation of the GWs, and
$e^A_{ab}(\hat\Omega)$ are the GW polarization tensors, which are uniquely defined by specifying $\hat m$ and $\hat n$ -- the GW principal axes:

\bea
e^+_{ab}(\hat\Omega) &=& \hat m_{a}\hat m_{b} - \hat n_{a}\hat n_{b} \, , \nonumber \\
 e^\times_{ab}(\hat\Omega) &=& \hat m_{a}\hat n_{b} + \hat n_{a}\hat m_{b}\, .
\eea

We note that General Relativity only predicts only two independent polarizations,
plus $+$, and cross $\times$, while other theories predict additional polarizations, such as breathing modes \cite{ljp08, ss12, grt15, IsiStein:2018}. Here we restrict ourselves to the well-known tensor transverse polarizations, $A=+,\times$, and refer the reader to e.g \cite{TeVeS:2018, cbc:2017, NG12p5AltPol} for an overview of alternative GW polarizations, and how they manifest in the GWB.

\begin{widetext}
We now have the ingredients to compute the energy density in GWs, and using the fact that $\dot{h}^{ab}=(\dot{h}^{ab})^*$:
\begin{equation}
	\label{eq:hdot}
	\left< \dot{h}_{ab}\dot{h}^{ab}\right> =
	\sum_A\sum_{A'} \! \int_{-\infty}^{\infty}\! \! \!\dif
	f \! \! \int_{-\infty}^{\infty}\! \! \!\dif
	f' \! \! \int_{S^2}\! \! \!\dif\hat\Omega  \!\! \int_{S^2} \! \! \! \dif\hat\Omega' \langle h_A(f,\hat\Omega)h^*_{A'}(f',\hat\Omega')\rangle
	e^A_{ab}(\hat\Omega)e^{ab}_{A'}(\hat\Omega') 4\pi^2 ff' \exp{[2\pi i(f\!-\!f')t-2\pi i(\hat\Omega-\hat\Omega')\cdot\vec{x}]}.
\end{equation}
\end{widetext}
For an isotropic, stationary, unpolarized, Gaussian stochastic
background, the quadratic expectation value of the Fourier modes is
given by \cite{Flanagan:1993}

\begin{equation} \label{eq:fourier-variance}
	\langle h_A(f,\hat\Omega)h^*_{A'}(f',\hat\Omega')\rangle = \frac{\delta_{AA'}}{2}\frac{\delta^{(2)}(\hat\Omega,\hat\Omega')}{4\pi}\frac{\delta(f-f')}{2}S_h(f),
\end{equation}
where $S_h(f)$ is the one-sided power spectral density (PSD) of the
Fourier modes of the GWB. This can also be written in terms of $H(f) = S_h(f)/(16 \pi)$,

\begin{equation} \label{eq:fourier-varianceH}
	\langle h_A(f,\hat\Omega)h^*_{A'}(f',\hat\Omega')\rangle = \delta_{AA'}\delta^{(2)}(\hat\Omega,\hat\Omega')\delta(f-f')H(f)\, ,
\end{equation}
and here $\langle \rangle$ denotes the ensemble average. Note that additional GW polarizations would also require modification of $H(f)$, e.g. \cite{ligo18}.

 Using this definition, together with
$\sum_Ae^A_{ab}e^{ab}_A = 4$, $\int\dif\hat\Omega=4\pi$ and Eqs. \eqref{eq:hdot} and \eqref{eq:fourier-variance},  we find
upon converting the frequency integral to $[0,\infty]$ that

\begin{equation}
	\langle\dot{h}_{ab}\dot{h}^{ab}\rangle = 8 \pi^2\int_0^{\infty}\dif f\; f^2S_h(f).
\end{equation}

Hence,
\begin{equation}
    \boxed{
	\Omega_\mathrm{gw}(f) \equiv \frac{1}{\rho_c}\frac{\dif\rho_\mathrm{gw}}{\dif\ln f} = \frac{2\pi^2}{3H_0^2}f^3S_h(f)}\, .\label{eq:omegaSh}
\end{equation}

 $\Omega_{\mathrm{gw}}(f)$
is also commonly reported in terms of $H(f)$, cf. Equation \eqref{eq:fourier-varianceH},

\beq
	\Omega_\mathrm{gw}(f)=\frac{32\pi^3}{3H_0^2}f^3 H(f) \, ,
\eeq
see e.g. \cite{AllenOttewill:1997, ar99, Anholm:2009, Mingarelli:2013, TaylorGair:2013,Moore2015}

\section{Power spectral density of the strain signal} \label{sec:psdStrain}
We now consider the strain signal in a GW experiment where the data-stream in the $i^{\rm th}$ detector, $s(t)$, will consist of a signal $h(t)$ and noise
$n(t)$,
\begin{equation}
s_i(t) = h_i(t) + n_i(t)\, .
\end{equation}

The one-sided PSD of the strain signal $S_{\mathrm{gw}}(f)$ is defined by

\begin{equation} \label{eq:strain-signal-psd}
	\langle \tilde{h}_i(f)\tilde{h}_j(f')\rangle = \frac{1}{2}\delta(f-f')S_{\mathrm{gw}}(f)_{ij},
\end{equation}
where tilde denotes a Fourier transform with the following convention:
\begin{equation}
	\tilde{h}_i(f) = \int_{-\infty}^{\infty} e^{-2\pi ift}h_i(t)dt.
\end{equation}

We can now explicitly evaluate the strain signal,
\begin{widetext}
\begin{equation}
	\tilde{h}_i(f) = \sum_A
	\int_{S^2}\dif\hat\Omega \int_{-\infty}^{\infty} \dif 
	f'  \int_{-\infty}^{\infty} \dif t e^{-2\pi
  	ift}F^A_i(\hat\Omega, f)h_A(f',\hat\Omega)e^{2\pi
  	if'(t-\hat\Omega\cdot\vec{x}_i)}
\end{equation}
\end{widetext}
where $F^A_i(\hat\Omega, f)$ is the antenna beam pattern response of the detector, see e.g. \cite{Mingarelli:2013,TaylorGair:2013,Mingarelli2014,2016ApJ...817...70T}.

The factor of $e^{2\pi ift}$ and its conjugate are such that the
integrals over frequency and time correspond to a forward and backward
Fourier transform, which simply leaves

\begin{equation}
	\tilde{h}_i(f) =
	\sum_A\int_{S^2}\dif\hat\Omega\;F^A_i(\hat\Omega, f)h_A(f,\hat\Omega)e^{-2\pi
  	if\hat\Omega\cdot\vec{x}_i}.
\end{equation}

Hence,
\begin{widetext}
\begin{equation}
	\langle \tilde{h}_i(f)\tilde{h}^*_j(f')\rangle =
	\sum_A\sum_{A'}\int_{S^2}\dif\hat\Omega\int_{S^2}\dif\hat\Omega'\langle
	h_A(f,\hat\Omega)h_{A'}^*(f',\hat\Omega')\rangle F^A_i(\hat\Omega, f)F^{A'}_j(\hat\Omega', f')e^{2\pi
  	i(f'\hat\Omega'\cdot\vec{x}_j-f\hat\Omega\cdot\vec{x}_i)}\, ,
\end{equation}
\end{widetext}
and using Equation\ \ref{eq:fourier-variance} this becomes
\begin{eqnarray}
	&&\langle \tilde{h}_i(f)\tilde{h}^*_j(f')\rangle = \frac{1}{16\pi}S_h(f)\delta(f-f') \nonumber \\
	&&\times \sum_A\int_{S^2}\dif\hat\Omega F^A_i(\hat\Omega, f)F^{A'}_j(\hat\Omega', f')e^{2\pi
	  if\hat\Omega \cdot(\vec{x}_j-\vec{x}_i)}\, .
\end{eqnarray}
Referring back to Equation\ \ref{eq:strain-signal-psd} we see that
\begin{equation} \label{eq:signal-mode-psd-compare}
S_{\mathrm{gw}}(f)_{ij} = \!\! \frac{S_h(f)}{8\pi} \! \sum_A\int_{S^2} \!\!\!  \dif\hat\Omega F^A_i(\hat\Omega, f)F^{A'}_j(\hat\Omega', f')e^{2\pi
  if\hat\Omega\cdot(\vec{x}_j-\vec{x}_i)}.
\end{equation}
The term which multiplies $S_h(f)/8\pi$ in Equation\ \ref{eq:signal-mode-psd-compare} is the un-normalized overlap reduction
function, which was first introduced in closed form in
\cite{Flanagan:1993}.

\subsection{Ground-based interferometers}
For co-located co-oriented interferometers this overlap
reduction function has a value of $8\pi/5$
\cite{Flanagan:1993, Allen:1996, AllenRomano:1999}. This number comes from the evaluation of the summation and integral term in Equation \ref{eq:signal-mode-psd-compare}, which does not include the factor of $8\pi$ in the denominator. Hence the PSD of the strain signal
in a given interferometer is related to the PSD of the Fourier modes
constituting the background via
\begin{align}
S_h(f) &= S_{\mathrm{gw}}(f)\times8\pi\times\frac{5}{8\pi} \nonumber\\
&= 5S_{\mathrm{gw}}(f).
\end{align}

It is important to distinguish between these two quantities when
computing the limit (or eventual detected value) of $\Omega_\mathrm{gw}(f)$, since the numerical factor of $5$ is unique to
ground-based interferometers, such as LIGO.

As an example of some confusion, we refer to Sec. 3.6 of
\citet{SathyaSchutz:2009}, where the fractional energy density in GWs
is defined as
\begin{equation}
\label{eq:sgw_ligo}
\boxed{\Omega_\mathrm{gw}(f) = \frac{10\pi^2}{3H_0^2}f^3S_{\mathrm{gw}}(f)}.
\end{equation}
The quantity $S_{\mathrm{gw}}(f)$ is described as the mean square
amplitude of the GW field per unit frequency, but it's important to
note that it is actually the PSD of the strain-signal in a single
ground-based interferometer, and not the PSD of the Fourier modes of the
GWB. The correct detector-independent definition of
$\Omega_\mathrm{gw}(f)$ is given by Equation\ \ref{eq:omegaSh}.

\subsection{Space-based gravitational wave detectors}
It is also possible to write $\Omega_\mathrm{gw}(f)$ in terms of $S_{\mathrm{gw}}(f)$ for space-based GW detectors such as LISA \cite{KudohEtAl:2006, Cornish:2001,Danzmann:1996}. For LISA, the overlap reduction function is normalized by $2/(5\sin^2\beta)$, 
where $\beta$ is the angle between the interferometer arms \cite{Cornish:2001}. For a LISA-type space-based GW detector, the proposed configuration of the arms is an equilateral triangle, however the effective angle $\beta=90\degree$. This is also called the A,E,T orthogonal signal configuration, see e.g.\cite{Adams_2010}, and when $\beta=60\degree$ this is called the X,Y,Z configuration. To be clear, we use the former. The PSD of the LISA strain signal is therefore related to the PSD of the Fourier modes of the GWB via
\begin{align}
S_h(f) &= S_{\mathrm{gw}}(f)\times8\pi\times\frac{5}{8\pi\sin^2\beta} \nonumber\\
&= \frac{5}{\sin^2\beta}S_{\mathrm{gw}}(f)\, , \\
&= 5\, S_{\mathrm{gw}}(f)\, ,
\end{align}
when $\beta= 90\degree$.

\subsection{Pulsar Timing Arrays}
Equation \eqref{eq:signal-mode-psd-compare} is also applicable to very low frequency GWs, which are
detectable by PTA experiments \cite{vanHaasterenEtAl:2011, DemorestEtAl:2013, Shannon:2013}.
Here, the value of the overlap reduction function for collocated and co-oriented pulsars is $8\pi/3$~(see e.g. \citep{Hellings1983}, and Appendix C of \cite{GairEtAl:2014}). Therefore the PSD
of the strain signal
in a given pulsar is related to the PSD of the Fourier modes
constituting the background via
\begin{align}
S_h(f) &= S_{\mathrm{gw}}(f)\times8\pi\times\frac{3}{8\pi} \nonumber\\
&= 3S_{\mathrm{gw}}(f) \, ,
\end{align}
although $S_{\mathrm{gw}}(f)$ is seldom reported in the PTA literature. Instead, PTAs report the detector-independent value
of $\Omega_\mathrm{gw}(f)$, Equation\ \ref{eq:omegaSh}, which can also be written in terms of the GW characteristic strain $h_c$,
where
\beq
\label{eq:hc}
h_c^2=f S_h(f) \, ,
\eeq
such that
\beq
\boxed{\Omega_\mathrm{gw}(f) = \frac{2\pi^2}{3H_0^2}f^2 h^2_c} \, .
\eeq

The characteristic strain can in turn be written as a function a dimensionless amplitude $A$ reported a reference frequency of $f_{yr}=1/\mathrm{yr}$:
\beq
\label{eq:hcA}
h_c = A \left(\frac{f}{f_{yr}}\right)^{\alpha} \, ,
\eeq
where $\alpha=-2/3$ for a stochastic GWB generated from the cosmic population of supermassive black hole binaries~\cite{Phinney:2001}. Using Equation \eqref{eq:hcA}, one can then
write down the expression for $\Omega_\mathrm{gw}$ found in \cite{Shannon:2013}:
\beq
\Omega_\mathrm{gw}(f)=\frac{2\pi^2}{3H_0^2} A^2 f^2_{yr}  \left(\frac{f}{f_{yr}}\right)^{2/3} \, .
\eeq

\section{Integrated bounds}
There are also indirect constraints on $\Omega_\mathrm{gw}$ from Cosmic Microwave Background temperature and polarization power spectra, from lensing, baryon acoustic oscillation, and Big Bang Nucleosynthesis, e.g. \cite{Maggiore:2000, spk06, ss:12, lms+16}. These constraints are integrated in frequency, and therefore are not directly comparable to $\Omega_\mathrm{gw}(f)$ limits, even though they are often plotted in the same figure. These bounds may be directly compared to other limits on $\Omega_\mathrm{gw}(f)$ by using power-law integrated curves derived in \cite{tr13}, and also applied and discussed in \cite{lms+16}.

\section{Scaling $\Omega_\mathrm{gw}(f)$ with time}

In order to understand how our sensitivity to $\Omega_\mathrm{gw}(f)$ increases
in time, we compute the maximum-likelihood estimator of the GW PSD. An independent derivation of the GWB scaling law for PTAs can be found in \cite{Siemens2013} and \cite{ar99} did this for LIGO, but here we do this in a detector-independent way. 

Consider a uniform-in-time sampling of noisy observations of GW strain; in the Fourier domain, the frequency spacing is $\Delta f=1/T$,
where $T$ is the length of the observation; if $\Delta F =
f_\mathrm{max}-f_\mathrm{min}$ is the bandwidth of the measurement, then the
number of independent frequency bins is $N=\Delta F/ \Delta f = T \Delta F$.
(If the full range of data are used, $f_\mathrm{max}$ is the Nyquist sampling
frequency and $f_\mathrm{min} = 0$ is the DC frequency; but in many cases only a
smaller bandwidth is informative or contains a stochastic signal.)

Under the assumption that the noise in the observations is stationary and
Gaussian, we can write the log-likelihood of the data in the frequency domain,
$s(f)$, as
\begin{multline}
\log(\mathcal{L}) \propto
-2 \Delta f \sum_{f \geq 0} \frac{ \left| s(f) \right|^2 }{[S_n(f)+S_{gw}(f)]} \\ - \sum_{f \geq 0} \log \left[ S_n(f) + S_{gw}(f)\right].
\end{multline}
(Recall that $S_{gw}(f)$ is the one-sided PSD; the sums above run over positive
frequencies.)

Optimizing the likelihood with respect to $S_{gw}(f)$, we obtain the
maximum-likelihood estimator for $S_{gw}(f)$, $\hat{S}_{gw}(f)$ in each
frequency bin:
\begin{equation}
  \hat{S}_{gw}(f) = 2 \Delta f \left| s(f) \right|^2 - S_n(f).
\end{equation}
The estimator is unbiased: the sampling mean, $\left \langle \hat{S}_{gw}(f)
\right\rangle = S_{gw}(f)$. The sampling variance of $\hat{S}$ is
\begin{equation}
  \var \hat{S}(f) = 2 \left( S_{gw}(f) + S_n(f) \right)^2,
\end{equation}
so the \emph{per-bin} S/N is
\begin{equation}
  \rho(f) = \frac{1}{\sqrt{2} \left( 1 + \frac{S_n(f)}{S_{gw}(f)}\right)}.
\end{equation}
Note that the per-bin S/N is independent of the observation time (additional
observation time leads to \emph{more bins}---finer frequency resolution---but
does not improve the uncertainty in any individual bin) and $0 \leq \rho(f) \leq
1/\sqrt{2}$, with the upper limit obtaining when $S_{gw}(f) \gg S_n(f)$.
Per-bin estimates of the PSD are never in the high-S/N limit \citep{Maggiore2007}.

Often we are not interested in per-bin estimates of the PSD, but instead want to
estimate the \emph{integrated} GW power over some bandwidth.  By
linearity, the (unbiased) estimator is just the integral of the estimator at
each frequency:
\begin{multline}
  \widehat{\int \mathrm{d} f \, S_{gw} (f)} \simeq \Delta f \sum_{f} \hat{S}_{gw}(f) = \\ \Delta f \sum_{f} \left[ 2 \Delta f \left| s(f) \right|^2 - S_n(f) \right].
\end{multline}
The sampling variance of this estimator is
\begin{equation}
  \Delta f^2 \sum_{f} 2 \left[ S_{gw}(f) + S_n(f) \right]^2,
\end{equation}
so the S/N of the integral estimate is
\begin{equation}
  \label{eq:integral-snr}
  \rho = \frac{\sum_{f} S_{gw} (f)}{\sqrt{\sum_{f} 2 \left( S_{gw}(f) + S_n(f) \right)^2}}.
\end{equation}
For a fixed bandwidth, $\Delta F = f_\mathrm{max} - f_\mathrm{min}$, the S/N in
Equation\ \eqref{eq:integral-snr} scales with the number of bins as $\sqrt{N} =
\sqrt{\Delta F / \Delta f} \propto \sqrt{T}$.  If the bandwidth \emph{also}
scales with time, as in the case for PTAs \citep{Siemens2013}, then the S/N can scale in a different way
than above.  Combining data from multiple independent measurements does not
change the scalings derived here; nor does working with unevenly sampled time
series, though care must be taken to define the effective bandwidth and
frequency resolution in this case.

\section{Discussion}
\label{disc}
We have derived $\Omega_\mathrm{gw}(f)$ in a detector-independent way, Equation \eqref{eq:omegaSh}, in the hopes that this will yield some clarity as to which expression is general (the former, written in terms of $S_h(f)$), and which expressions are detector-dependent, e.g. Equation \eqref{eq:sgw_ligo}.

Next-generation GW observatories, such as Cosmic Explorer and the Einstein Telescope, will enable the detection of an unprecedented number of compact binary coalescences, generating a significant astrophysical GWB at high frequencies \cite{Regimbau:2016ike} that poses a challenge for detecting weaker cosmological signals \cite{Sachdev:2020bkk, Zhou:2022nmt}. This astrophysical GWB arises from both undetected mergers and residual foreground contamination from imperfectly subtracted resolved signals, primarily due to uncertainties in the inference of key parameters such as coalescence phase and luminosity distance \cite{Zhou:2022nmt}. A major obstacle in searching for cosmological GWBs is thus not only the presence of this astrophysical background but also the limitations of existing subtraction techniques. Time-frequency domain notching has been explored as a strategy to mitigate the foreground, suppressing it to approximately 5\% of its original level, but the remaining unresolvable binary neutron star background remains dominant and requires improved detector sensitivity to address \cite{Zhong:2024dss}.

In addition to the stochastic background, overlapping signals introduce ``source confusion,'' \cite{Janquart:2022fzz, Pizzati:2021apa,Relton:2022whr} where multiple compact binary signals simultaneously exist in the detector's frequency band. While neutron star binaries, given their long inspiral duration, frequently overlap in the time-frequency domain, particularly at low frequencies below 5 Hz, their impact on parameter estimation is minimal, as significant confusion only arises when signals have nearly identical chirp masses and overlap at frequencies above 40 Hz are a rare occurrence \cite{Johnson:2024foj}. Similarly, for short-duration binary black hole mergers, foreground noise from overlapping signals can alter the noise power spectrum and reduce detection sensitivity by approximately 25\%, but standard parameter estimation methods remain effective without requiring complex global-fit techniques or signal subtraction \cite{Gupta:2024lft}. Therefore, while the astrophysical GWB constitutes a fundamental limit to the detection of cosmological GWBs, source confusion and foreground noise minimally impact individual event parameter inference. To enhance sensitivity to a cosmological background, future efforts must focus on refining signal subtraction techniques and improving detector capabilities to better suppress the dominant astrophysical foreground.

To summarize, the measured value of $\Omega_\mathrm{gw}(f)$ from different GW detectors does not need to be rescaled -- the overlap reduction functions are already folded into the final expressions. Here we showed how these expressions are related for LIGO, LISA and PTA experiments, in a general framework, so that one may also carry out this calculation with ease for future GW detectors. We hope that the calculations carried out here are comprehensive, and bring some more clarity to how physical this physical quantity can be measured in the new and emerging field of GW astrophysics and cosmology. \\

\section*{Software}
\noindent Figure 1 can be reproduced using our open access code on GitHub, \url{https://github.com/ChiaraMingarelli/omega_gw}.

\section*{Acknowledgments}
The authors thank David Spergel, Thomas Callister, Tristan Smith, Joseph Romano, Neil Cornish, Bjorn Larsen, and Qinyuan Zheng for their comments on this manuscript. We also thank Christopher Moore for valuable conversations. The Flatiron Institute is supported by the Simons Foundation. C.M.F.M receives support from NSF grant AST-2106552, and C.M.F.M. and S.R.T acknowledge support from the NANOGrav NSF Physics Frontier Center award numbers PHY-1430284 and PHY-2020265. SRT also acknowledges support from AST-2007993, and an NSF CAREER Award PHY-2146016. B.S.S is supported in part by NSF grants AST-2307147, PHY-2207638, PHY-2308886 and PHY-2309064. This research was also supported in part by the NSF grant PHY-1748958.
\vspace{0.1cm}
\bibliography{pta}
\end{document}